\documentclass[twocolumn]{aastex63}
\usepackage[utf8]{inputenc}

\shorttitle{Diamagnetic-blob accretion in TX Col}
\shortauthors{Littlefield et al.}

\newcommand{\tess}{\textit{TESS}}

\begin{document}

\title{Quasi-periodic oscillations in the \tess\ light curve of TX Col,\\a diskless intermediate polar on the precipice of forming an accretion disk}

\author{Colin Littlefield}
\affiliation{University of Notre Dame, Notre Dame, IN 46556, USA}
\affiliation{Department of Astronomy, University of Washington, Seattle, WA 98195, USA}

\author{Simone Scaringi}
\affiliation{Centre for Extragalactic Astronomy, Department of Physics, University of Durham, South Road, Durham DH1 3LE, UK}

\author{Peter Garnavich}
\affiliation{University of Notre Dame, Notre Dame, IN 46556, USA}

\author{Paula Szkody}
\affiliation{Department of Astronomy, University of Washington, Seattle, WA 98195, USA}

\author{Mark R. Kennedy}
\affiliation{Jodrell Bank Centre for Astrophysics, School of Physics and Astronomy, The University of Manchester, Manchester M13 9P, UK}

\author{Krystian I{\l}kiewicz}
\affiliation{Centre for Extragalactic Astronomy, Department of Physics, University of Durham, South Road, Durham DH1 3LE, UK}
\affiliation{Department of Physics and Astronomy, Texas Tech University, PO Box 41051, Lubbock, TX 79409, USA}

\author{Paul A. Mason}
\affiliation{New Mexico State University, MSC 3DA, Las Cruces, NM, 88003, USA}
\affiliation{Picture Rocks Observatory, 1025 S. Solano Dr. Suite D, Las Cruces, NM 88001, USA}

\submitjournal{AJ; accepted for publication}
\received{April 29, 2021}
\revised{May 22, 2021}
\accepted{May 26, 2021}

\correspondingauthor{Colin Littlefield}
\email{clittlef@alumni.nd.edu}

\begin{abstract}

    One of the fundamental properties of an intermediate polar is the dynamical nature of the accretion flow as it encounters the white dwarf's magnetosphere. Many works have presumed a dichotomy between disk-fed accretion, in which the WD accretes from a Keplerian disk, and stream-fed accretion, in which the matter stream from the donor star directly impacts the WD's magnetosphere without forming a disk. However, there is also a third, poorly understood regime in which the accretion flow consists of a torus of diamagnetic blobs that encircles the WD. This mode of accretion is expected to exist at mass-transfer rates below those observed during disk-fed accretion, but above those observed during pure stream-fed accretion. We invoke the diamagnetic-blob regime to explain the exceptional \tess\ light curve of the intermediate polar TX~Col, which transitioned into and out of states of enhanced accretion during Cycles 1 and 3. Power-spectral analysis reveals that the accretion was principally stream-fed. However, when the mass-transfer rate spiked, large-amplitude quasi-periodic oscillations (QPOs) abruptly appeared and dominated the light curve for weeks. The QPOs have two striking properties: they appear in a stream-fed geometry at elevated accretion rates, and they occur preferentially within a well-defined range of frequencies ($\sim$10-25~cycles~d$^{-1}$). We propose that during episodes of enhanced accretion, a torus of diamagnetic blobs forms near the binary's circularization radius and that the QPOs are beats between the white dwarf's spin frequency and unstable blob orbits within the WD's magnetosphere. We discuss how such a torus could be a critical step in producing an accretion disk in a formerly diskless system.

\end{abstract}


    \section{Introduction}

    In an intermediate polar (IP), a magnetized white dwarf (WD) accretes from a low-mass companion star that overfills its Roche lobe. The magnetic-field strength of the WD is strong enough to channel the accretion flow as it approaches the WD \citep[for a review, see][]{patterson94}. The defining property of IPs as a class is the asynchronous rotation of the WD, whose rotational frequency ($\omega$) is higher than the binary orbital frequency ($\Omega$). The inequality of $\omega$ and $\Omega$ is believed to be a stable equilibrium condition, as opposed to a short-lived deviation from synchronous rotation (i.e., where $\omega = \Omega$). 
    
    The dynamical nature of the accretion flow is one of the basic characteristics of any individual IP. There is widespread consensus in both the observational and theoretical literature that IPs can accrete from a Keplerian accretion disk or a ballistic accretion stream that directly impacts the magnetosphere. The former is usually referred to as ``disk-fed accretion,'' and the latter as ``stream-fed'' or ``diskless'' accretion. The mode of accretion in a particular IP can change in response to variations in the system's mass-transfer rate ($\dot{M}$). FO~Aqr, which shows disk-fed accretion when $\dot{M}$ is highest and some form of stream-fed accretion when $\dot{M}$ decreases \citep{littlefield}, is perhaps the best example of this phenomenon. Overall, disk-fed accretion tends to be the most commonly observed mode of accretion in the majority of IPs.
    
    Optical and X-ray power spectra provide a common means of distinguishing between the various modes of accretion \citep{fw99}. In general, disk-fed accretion is expected to result in a dominant photometric signal at $\omega$, because magnetically entrained material is lifted from an azimuthally uniform disk and is then forced to corotate with the WD. The resulting structure is known as an accretion curtain, and as it rotates, its changing aspect produces a photometric modulation. 
    
    In contrast, the stream-fed regime \citep{fw99} assumes that the WD is supplied by infalling matter from a fixed location within the binary rest frame. Therefore, power will appear at the spin-orbit beat frequency ($\omega-\Omega)$, which is the frequency at which the WD's magnetosphere (rotating at $\omega$) interacts with a fixed structure in the binary rest frame (revolving at $\Omega$). If a second accreting pole is visible, stream-fed accretion can also shift power to $2(\omega-\Omega)$, because the accretion flow will switch between poles due to the WD's asynchronous rotation.

    \citet{kw99} have pointed out that in spite of its name, ``diskless'' accretion does not require the absence of a disk-like structure. Instead, its essential property is simply that the accretion flow is non-Keplerian, and \citet{wk95} calculated that the flow can consist of short-lived diamagnetic blobs that enter unstable orbits in the WD's magnetosphere. However, they do not survive long enough to interact viscously to form an accretion disk \citep{wk95}. Consistent with these predictions, observations of the diskless intermediate polar V2400~Oph suggest that while the system lacks a Keplerian disk, the accretion flow nevertheless encircles the WD at all azimuths \citep{hellier02}. There are no clear theoretical predictions about the power spectrum of a system experiencing blob-fed accretion; this regime was not modelled in \citet{fw99} or in any other study of which we are aware.
    
    Finally, we note that very few observations of IPs by either the \textit{Kepler} spacecraft or the Transiting Exoplanet Survey Satellite (\textit{TESS}) have been published. FO~Aqr \citep{kennedy} and RZ~Leo \citep{rzleo} are two exceptions. MV~Lyr, whose magnetic field is so weak that it can channel the accretion flow only at very low accretion rates, is another such system \citep{scaringi17}.

    \subsection{TX Col}
    
    Originally discovered by \citet{tuohy86}, the intermediate polar TX~Col was the subject of a comprehensive study by \citet{bt89}, who identified orbital and spin periods of 5.7~h and 1911~sec, respectively. TX~Col is perhaps most notable as a compelling example of an IP that has alternated between various modes of accretion, based on the eclectic behavior of its power spectrum across different epochs. However, the details of accretion in TX Col have proven frustratingly difficult to pinpoint. Optical power spectra in 1984-1985 contained a dominant signal at $\omega-\Omega$, but in November 1989, the power shifted to $2(\omega-\Omega)$ \citep{buckley92}. Neither of these frequencies was present during January 1994, when low-frequency quasi-periodic oscillations (QPOs) near $\sim5000$~s dominated the light curve \citep{sullivan}. \citet{mhlahlo} found that the \citet{sullivan} QPOs have been detected at various epochs over the course of 12 years and hypothesized that it is a beat between diamagnetic blobs in the outer accretion disk and the WD spin period. 
    
    TX~Col's behavior has been equally enigmatic in X-rays. \citet{norton} reported X-ray observations from 1994 October and 1995 October and found via power-spectral analysis that the WD was accreting from a disk in the first observation and from a combination of a disk and stream one year later. The unobserved transition between these states meant that identifying a unique explanation for the transition, such as a changed mass-transfer rate, was not possible.
    
    No clear consensus has emerged about how to best interpret the nature of TX~Col's accretion flow. Indeed, in their theoretical examination of the optical power spectra of IPs, \citet{fw99} specifically pointed to TX~Col as a challenge to their model. Although the presence of $\omega-\Omega$ and $2(\omega-\Omega)$ is consistent with stream-fed accretion, optical spectra of TX~Col do not show the high infall velocities expected of stream-fed accretion.
    
    The Gaia EDR3 distance of TX~Col is 909$^{+18}_{-21}$~pc \citep{BJ21}.

    \begin{figure*}
        \centering
        \includegraphics[width=\textwidth]{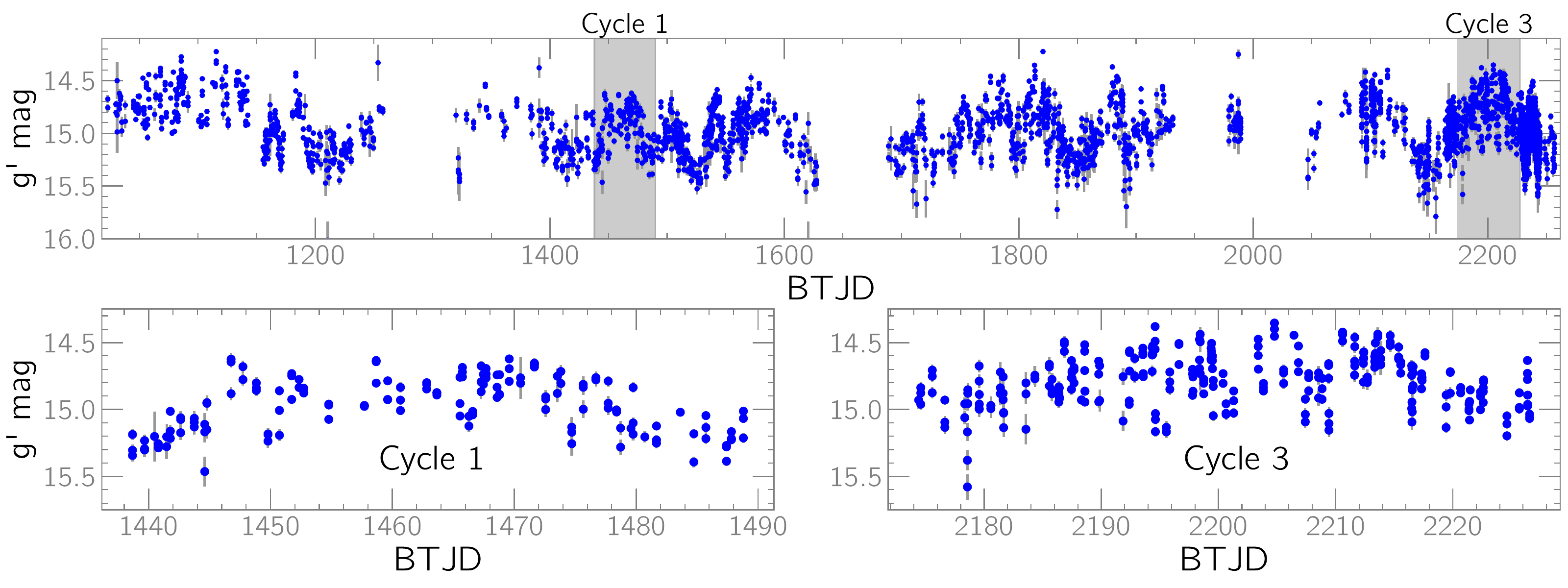}
        \caption{{\bf Top panel:} The full ASAS-SN $g'$-band light curve. The shaded regions indicate the epochs of the \tess\ observations of TX Col during Cycles 1 and 3. BTJD is equivalent to BJD + 2457000. The accretion states observed by \tess\ have been common in TX~Col's ASAS-SN light curve since BTJD$\sim$1150. {\bf Bottom panels:} Enlargements of the ASAS-SN data obtained during the \tess\ observations in Cycle 1 (left) and Cycle 3 (right). The bright state during Cycle 1 was 0.6~mag brighter than the normal state. During Cycle 3, the bright state was comparatively amorphous, characterized by a slow brightening of several tenths of a magnitude, followed by a gradual dimming. The ASAS-SN observations confirm that the bright states observed in the \tess\ SAP light curve are of astrophysical origin.
        }
        \label{fig:ASAS-SN}
    \end{figure*}

    \section{Data}
    
    The \textit{TESS} spacecraft observed TX Col at a two-minute cadence between 2018 November 15 and 2019 January 6 during Cycle 1 and between 2020 November 20 and 2021 January 13 in Cycle 3. We extracted TX~Col's simple-aperture-photometry (SAP) flux from each of these four sectors using {\tt lightkurve}  \citep{lightkurve}. The timestamps were expressed in the Barycentric Tess Julian Date (BTJD) time standard, which is defined as BTJD = BJD + 2457000, where BJD is the Barycentric Julian Date in Barycentric Dynamical Time.
 
    During final preparation of this manuscript, \citet{rawat} published their study of the Cycle~1 light curve. Their analysis differs from ours in that theirs relies upon the pipeline-created PDCSAP light curve, which attempts to correct systematic trends in a target's flux. However, the PDCSAP flux can suppress genuine astrophysical variability,\footnote{See Sec.~2.1 of the \tess\ Archive Manual at \url{ https://outerspace.stsci.edu/display/TESS/2.1+Levels+of+data+processing}. } particularly slow stochastic variations. This likely accounts for the significant differences between the SAP light curve presented here and the PDCSAP light curve analyzed in \citet{rawat}.
    
    TX~Col suffers from mild blending in the \textit{TESS} data. To quantify this blending, we relied upon the $g'$-band light curve from the All-Sky Survey for Supernovae \citep[ASAS-SN;][]{shappee, kochanek}, which includes 122 observations obtained simultaneously with a TESS image in Cycle 1 and 261 in Cycle 3. We plotted the \textit{TESS} flux as a function of the ASAS-SN flux and interpreted the $y$-intercept (\textit{i.e.}, the anticipated TESS count rate at an ASAS-SN flux of 0 mJy) as the unvarying, contaminating flux. We subtracted this level from the light curve. Because the orientation of the image was different in each sector, we performed this procedure separately for each sector in order to allow for the possibility of different levels of contamination. This method is undoubtedly an oversimplification in that it assumes that the variability in the \tess\ bandpass has a one-to-one correlation with $g$-band variability, but as we will see later, the two deblended light curves in each Cycle are consistent with each other.

    \section{Analysis}
    
    \subsection{Cycle 1 Light Curve}
    
    The Cycle~1 \textit{TESS} light curve is centered on a four-week-long bright state of TX~Col.\footnote{The bright state is present in both the ASAS-SN and the SAP light curves, but the PDCSAP light curve suppresses it.} According to contemporaneous ASAS-SN $g'$-band observations (Fig.~\ref{fig:ASAS-SN}), the bright state was only $\sim$0.6-0.7~mag above TX~Col's normal brightness. The long-term ASAS-SN light curve of TX~Col shows that bright states of this amplitude and duration have been common in the past three years.

\begin{figure*}
        \centering
        \includegraphics[width=\textwidth]{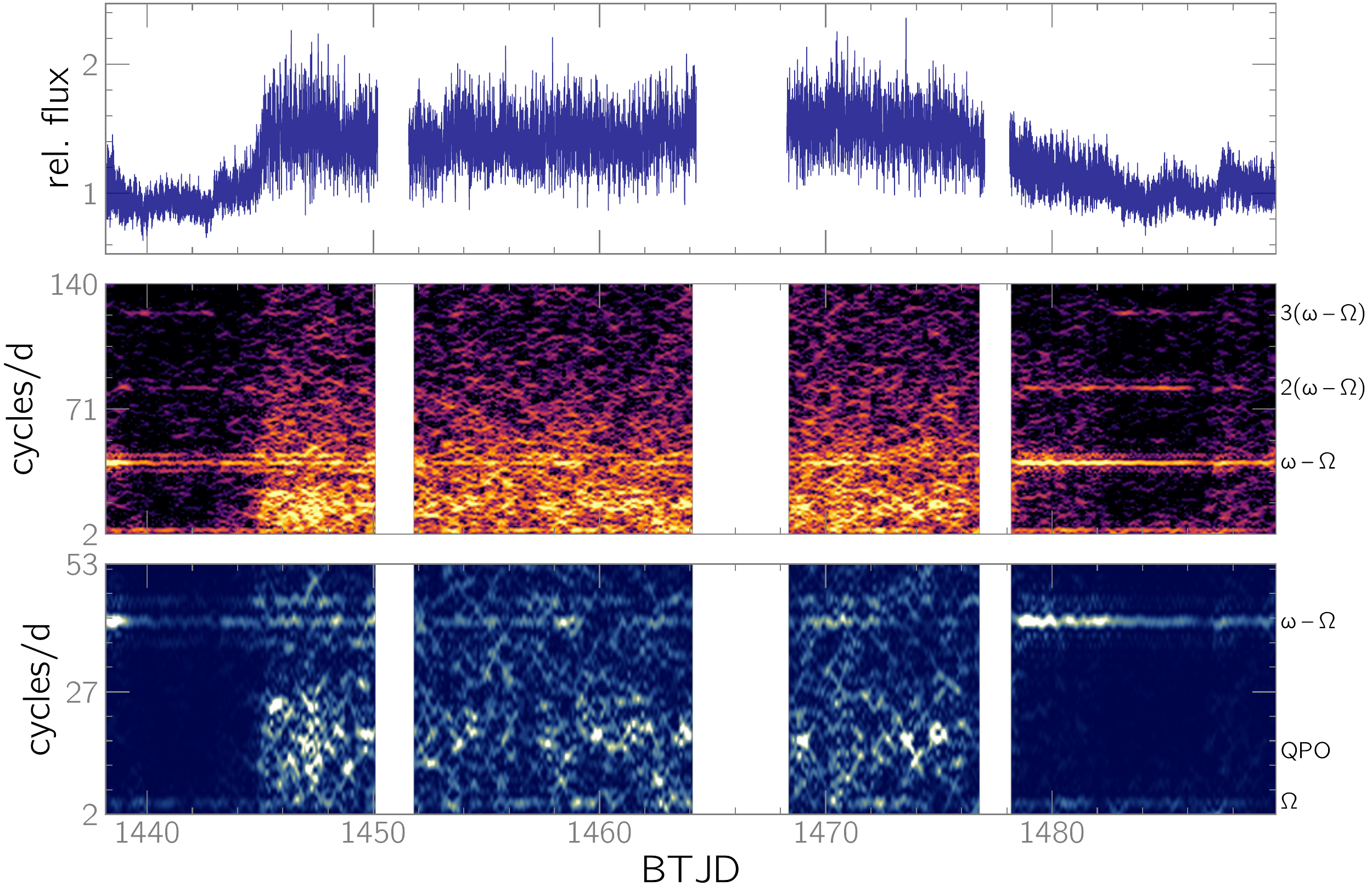}
        \caption{{\bf Top:} The Cycle 1 \tess\ light curve. The overall shape of the light curve correlates well with the contemporaneous ASAS-SN data (Fig.~\ref{fig:ASAS-SN}), confirming that the variations are of astrophysical origin. {\bf Middle:} Trailed power spectrum, with a logarithmic scaling. The constituent power spectra used a window size of 0.5~d and were not normalized. {\bf Bottom:} Trailed power spectrum, with a linear scaling. Major frequencies are identified at the right side of the trailed power spectrum in terms of the spin frequency $\omega$, the orbital frequency $\Omega$, and the beat frequency $\omega-\Omega$. Also shown is the QPO frequency reported by \citet{mhlahlo}. The QPOs appear only during the bright state and persist for four weeks.}
        \label{fig:trailed_power}
    \end{figure*}

\begin{figure*}
        \centering
        \includegraphics[width=\textwidth]{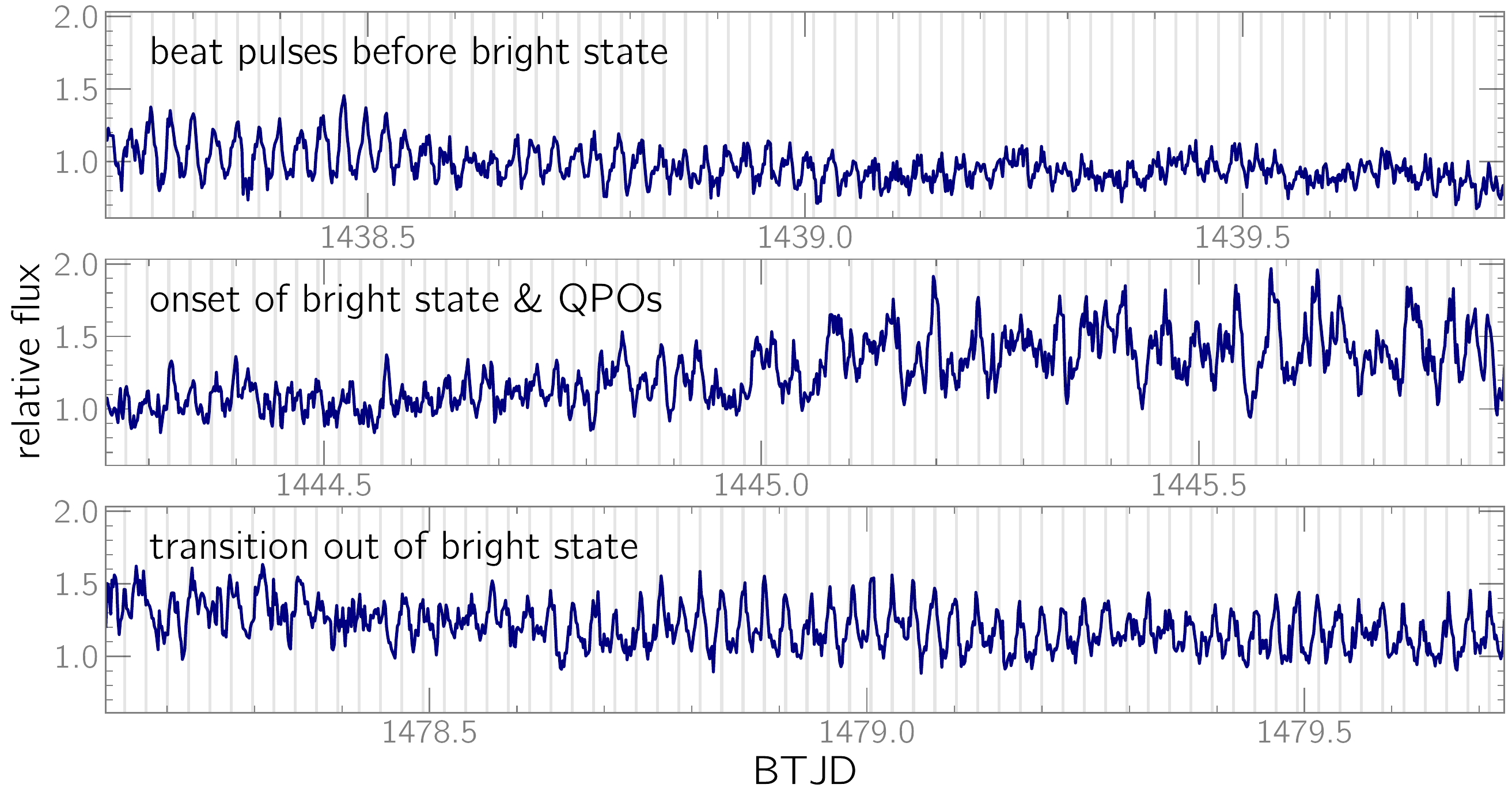}
        \caption{Three segments of the Cycle 1 light curve. The vertical lines indicated expected times of maximum light of the beat pulses. In the top and bottom panels, the beat pulse achieves its highest amplitude and sharpest profile when the system's brightness is declining.}
        \label{fig:lc_segments}
    \end{figure*}

    What makes the bright-state light curve extraordinary is the rapid evolution of the power spectrum as TX~Col brightened. This is best seen in the trailed power spectrum in Fig.~\ref{fig:trailed_power}. Before TX~Col entered its bright state, the light curve showed a coherent signal at the beat frequency ($\omega-\Omega$) and its harmonics, with comparatively little power at the spin frequency ($\omega$). But during the bright state, the power spectrum was fundamentally different, with a series of large-amplitude QPOs overwhelming the periodic variability in the two-dimensional power spectrum. In fact, the QPOs dominated the light curve for the entire four-week bright state (Fig.~\ref{fig:trailed_power}). Then, as TX~Col faded from its bright state, the QPOs vanished in the span of less than one binary orbit, with $\omega-\Omega$ reemerging just as abruptly (Fig.~\ref{fig:lc_segments}). Based on the simultaneous ASAS-SN observations, $g\sim15.0$ was the threshold brightness for the emergence and disappearance of the QPOs.

    Fig.~\ref{fig:trailed_power} shows QPOs across a wide range of frequencies during the bright state, but it is striking that the highest-amplitude QPOs occur within a well-defined frequency range. This is best seen in linearly scaled trailed power spectrum (Fig.~\ref{fig:trailed_power}, bottom panel), where the strongest QPOs appear in a corridor between $\sim10-25$ cycles d$^{-1}$.

\begin{figure}
        \centering
        \includegraphics[width=\columnwidth]{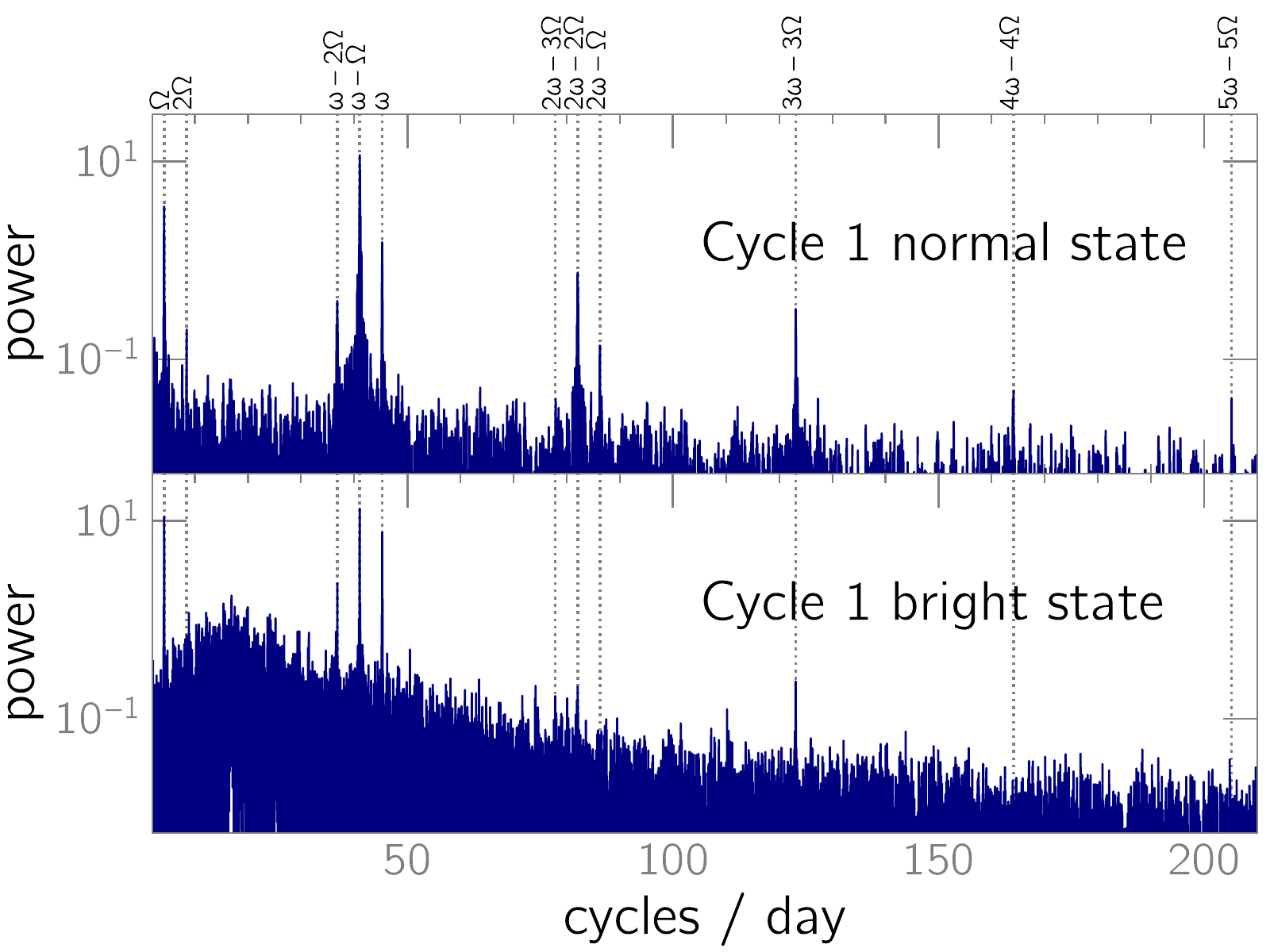}
        \caption{Power spectra of TX~Col during the two brightness states in Cycle~1. Major frequencies are identified relative to the spin frequency $\omega$ and the orbital frequency $\Omega$. The bump near $\sim$20~cycles d$^{-1}$ during the bright state is from quasi-periodic oscillations.}
        \label{fig:power}
    \end{figure}

    The one-dimensional power spectra of the two accretion states (Fig.~\ref{fig:power}) contain a similar ensemble of frequencies, dominated by $\omega-\Omega$ and its harmonics and sidebands. The major difference between the two power spectra is a precipitous increase in quasi-periodic variability that manifests itself as a wide bulge in the power spectrum near $\sim20$~cycles~d$^{-1}$ during the bright state.

    \begin{figure*}
        \centering
        \includegraphics[width=\textwidth]{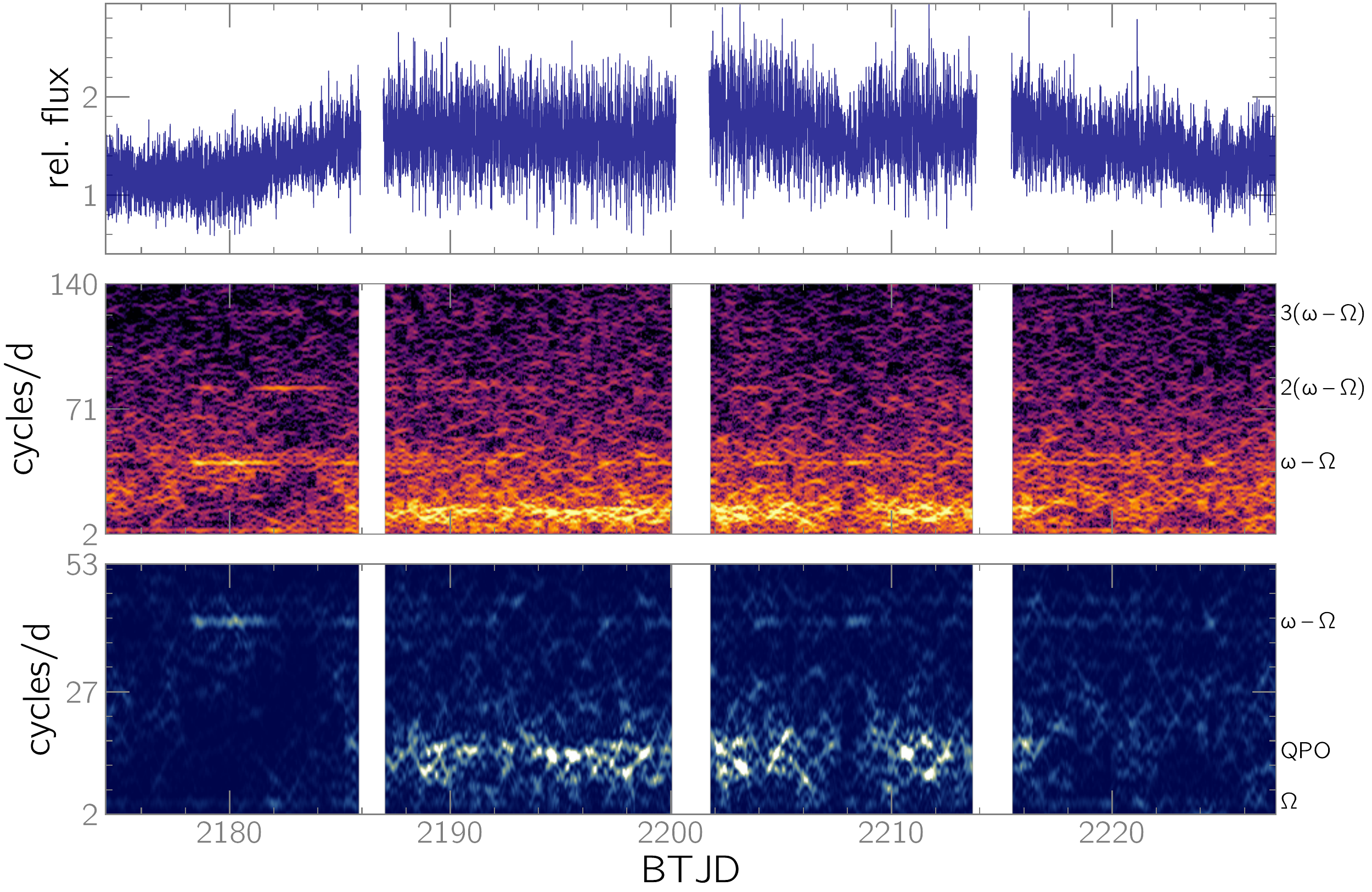}
        \caption{As with Fig.~\ref{fig:trailed_power}, but for the Cycle 3 data. The flux is relative to the quiescent level in Cycle 1. Compared to the Cycle 1 data, the dips during the bright state were deeper, and the power spectrum outside of the bright state generally lacked the well-defined, periodic variability seen during the corresponding intervals in Cycle 1.}
        \label{fig:trailed_power_cycle3}
    \end{figure*}

    \subsection{Cycle 3 Light Curve}
    
    \begin{figure}
        \centering
        \includegraphics[width=\columnwidth]{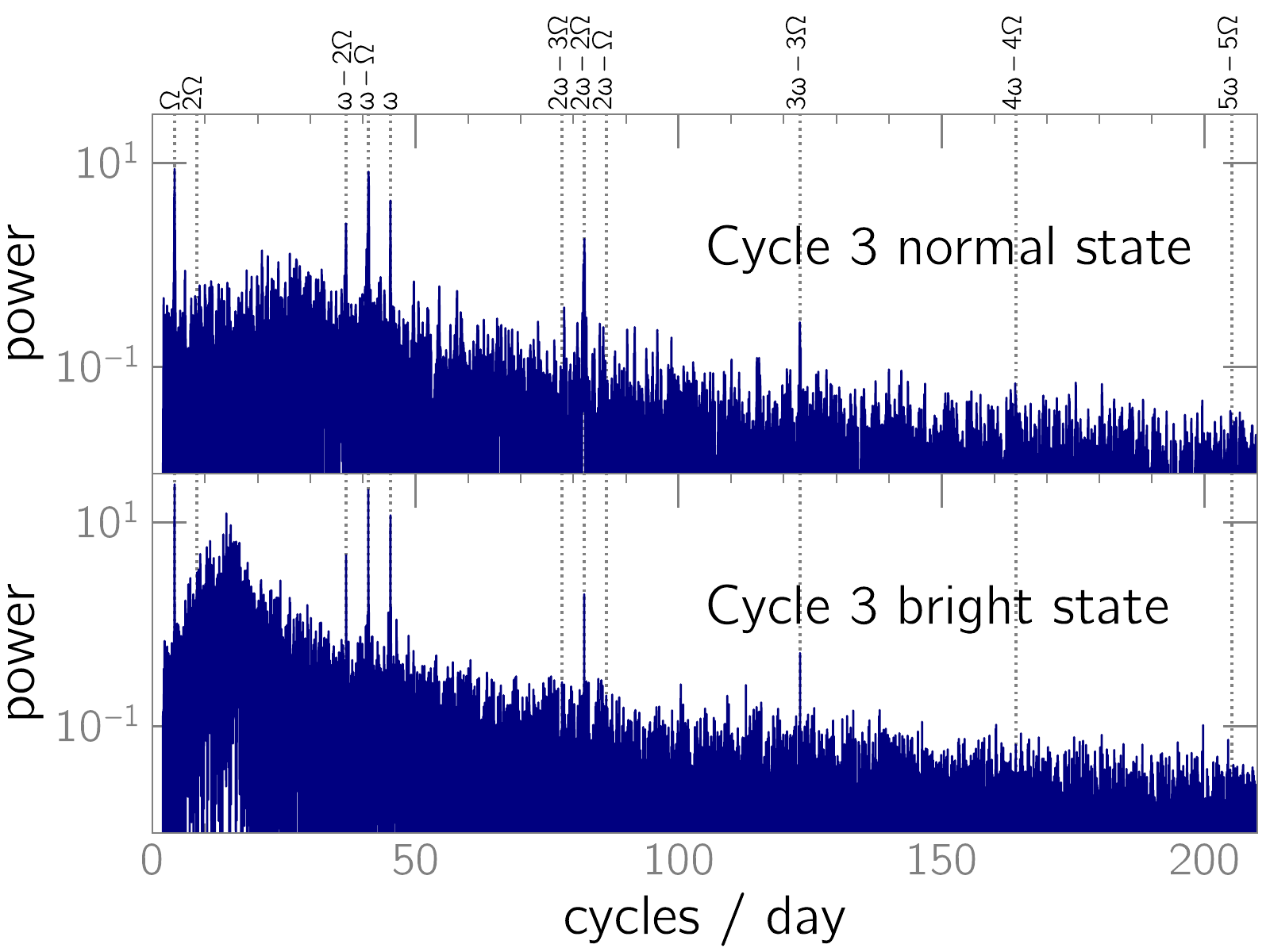}
        \caption{Power spectra of the two brightness states in Cycle 3. QPOs create an obvious bump near 20~cycles~d$^{-1}$.}
        \label{fig:cycle3_power}
    \end{figure}

    At first glance, the Cycle 3 light curve closely resembles the Cycle 1 data. There were no QPOs during the first two weeks of the Cycle 3 observation, when TX~Col was comparatively faint. As with the Cycle~1 data, they returned during the central four weeks as TX Col gradually brightened (Fig.~\ref{fig:trailed_power_cycle3}). 
    
    The one-dimensional power spectrum (Fig.~\ref{fig:cycle3_power}) establishes that the major harmonics and sidebands from the Cycle~1 power spectrum were present in Cycle~3, although the QPOs were even more prominent in Cycle~3. In neither cycle was there any evidence of the candidate superhump frequencies at 3.4~cycles~d$^{-1}$ and 4.8~cycles~d$^{-1}$ reported by \citet{retter}.
    
\begin{figure*}
        \centering
        \includegraphics[width=\textwidth]{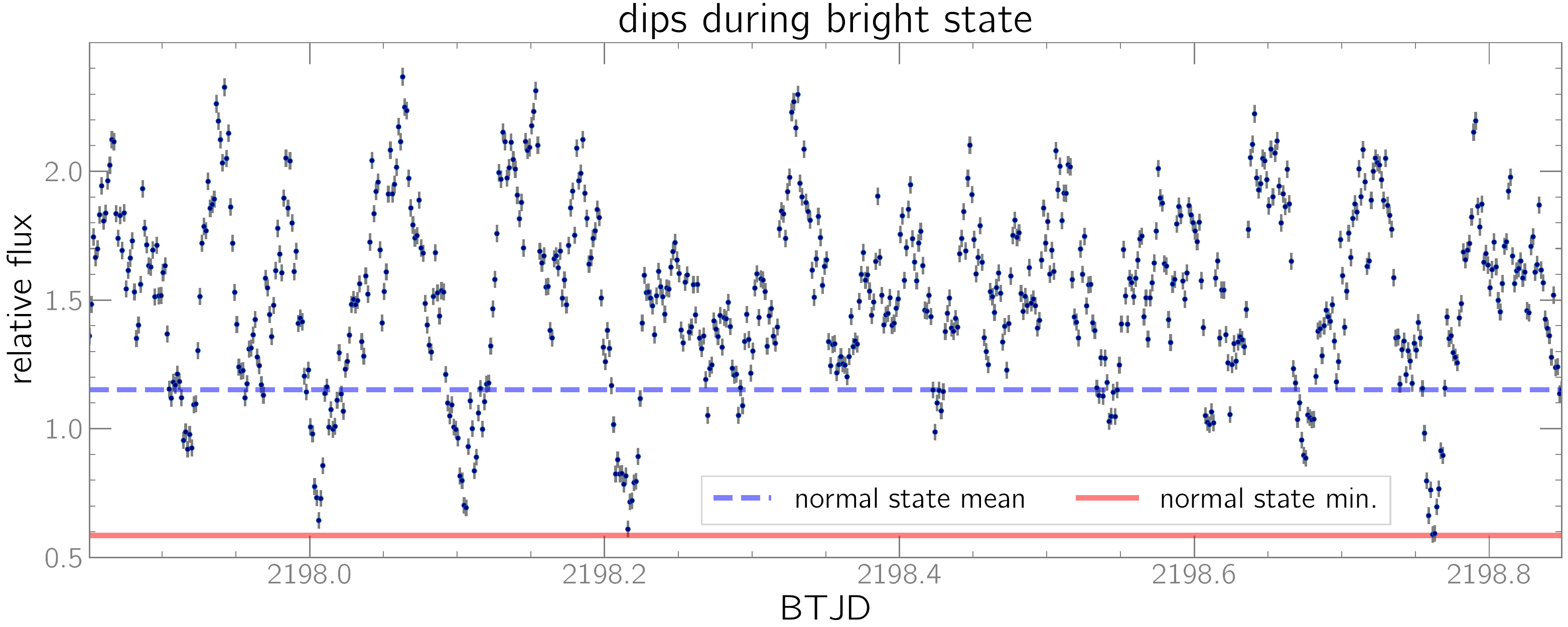}
        \caption{Representative section of the Cycle 3 bright state, showing stochastic dips. The dashed blue line and the solid red line indicate, respectively, the average brightness and minimum brightness before the bright state. }
        \label{fig:dips}
    \end{figure*}

    Overall, the behavior of the QPOs during Cycle~3 was comparable to that of the Cycle~1 QPOs. For example, when the bright state began in Cycle~3, large-amplitude QPOs appeared between $10-25$~cycles~d$^{-1}$, and at the same time, the amplitudes of  $\omega-\Omega$ and its harmonics were greatly suppressed. The QPOs were present throughout the bright state, except for a very brief interruption near BTJD=2209, when there was a temporary diminution of their amplitude. Additionally, the QPOs were interspersed with a series of stochastic dips that lasted for as long as $\sim$30~minutes, occasionally fading to the pre-bright-state level (Fig.~\ref{fig:dips}). Although dips were also present during the Cycle~1 QPOs, they tended to be shallower and less pronounced than their counterparts in Cycle~3. These dips are suggestive of brief interruptions of accretion onto the WD. Rapid, unexplained dips have been seen in other IPs during their outbursts (\textit{e.g.}, EX Hya, \citealt{ex_hya_dip}), but it is unclear whether this behavior is related to the dips during TX~Col's bright state.

    The behavior near the end of the Cycle~3 bright state is notably different than the end of the Cycle~1 bright state. In Cycle~1, there was a very clean dichotomy between its QPO-dominated bright state and its $\omega-\Omega$-dominated normal state, as is evident from the two-dimensional power spectrum in Fig.~\ref{fig:trailed_power}. Conversely, the end of the bright state in Cycle 3 was not as clearly defined in the power spectrum (Fig.~\ref{fig:trailed_power_cycle3}). Although large-amplitude QPOs ceased near BTJD=2215, lower-amplitude QPOs remained prominent in the power spectrum (Fig.~\ref{fig:trailed_power_cycle3}, central panel) and the amplitudes of the periodic signals, such as $\omega-\Omega$, did not increase. The poorly defined end of Cycle~3's bright state suggests that the bright-normal dichotomy from Cycle~1 is an oversimplification and that there is an intermediate brightness regime in which the QPO amplitudes are comparable to the amplitudes of the periodic variability.

    The contemporaneous ASAS-SN observations (Fig.~\ref{fig:ASAS-SN}) can elucidate some of the differences between the two TESS observations. During Cycle~1, the normal brightness level was $g'\sim15.3$, and the bright state peaked at $g'\sim14.8$. In contrast, the ASAS-SN data during Cycle~3 began at $g'\sim14.9$ and brightened by several tenths of a magnitude when the QPOs appeared. Hence, TX~Col was brighter at all times during Cycle~3, as compared to Cycle-1, indicating an increase in its accretion rate. This might explain why the end of the Cycle~3 bright state is ill-defined, both in the light curve and in the two-dimensional power spectrum.

    \section{Discussion}    \label{sec:discussion}

    \subsection{Mode of accretion}
    
    Theoretical modeling of IP power spectra by \citet{fw99} predicts that disk-fed accretion should produce a dominant signal at $\omega$, while stream-fed accretion will tend to shift power into $\omega-\Omega$ and $\Omega$. Moreover, \citet{fw99} find that $\omega$ can exist at a diminished amplitude even in stream-fed systems, particularly if the stream-magnetosphere interaction occurs across a wide range of azimuth. Thus, the mere presence of $\omega$ is not uniquely indicative of the presence of a disk, particularly when there is substantial power at $\omega-\Omega$, its harmonics, and its sidebands. For example, an amplitude modulation of the beat frequency across the orbit \citep{warner86} can shift power equally into the upper and lower orbital sidebands of the beat frequency---\textit{i.e.}, $(\omega - \Omega) \pm \Omega = \omega$ and $\omega - 2\Omega$.
    
    Based on this framework, we infer that TX~Col was consistently in a stream-fed geometry throughout the \tess\ observations and that no accretion disk was present. In particular, the power spectra establish that the amplitude of $\omega$ was consistently below that of $\omega-\Omega$, an observation that is challenging to explain in a disk-fed geometry but expected for a stream-fed IP. Moreover, the harmonics of $\omega$ in TX~Col's power spectrum were extremely weak, while those of $\omega-\Omega$ were pronounced. The conspicuous signal at $\omega-2\Omega$ is another important clue, because it suggests that $\omega-\Omega$ might be amplitude-modulated at $\Omega$, shifting power into $\omega$ and $\omega-2\Omega$ as predicted by \citet{warner86}. Not only do we observe significant power at both frequencies, but \citet{fw99} also predict that stream-fed accretion can cause the amplitude of $\omega-2\Omega$ to rival that of $\omega$, as observed in TX~Col. Thus, a stream-fed geometry offers a more natural explanation for the power spectrum of TX~Col.
    
    The QPOs present a significant challenge to applying \citet{fw99} to TX~Col. Their theoretical modeling of stream-fed accretion predicts periodic variability at $\omega-\Omega$ and its first harmonic; various sidebands, such as $2\omega-\Omega$, are also possible, but at a basic level, the variability is expected to be periodic if the WD accretes from a stationary region in the binary rest frame \citep{fw99}. TX~Col validates this prediction---but only outside of its bright state. The abrupt breakdown of its periodic photometric variability during the bright state suggests that there are at least two $\dot{M}$-dependent regimes of diskless accretion: one in which the observed variability is periodic (low $\dot{M}$) and another in which it is mostly quasi-periodic (higher $\dot{M}$).

    The stream-fed geometry before and after the bright state has several interesting implications. First, the absence of an accretion disk at the start of the bright state precludes the possibility that it was an outburst produced by the dwarf-nova instability. Second, the mechanism of the QPOs must be independent of the presence of a disk. From these inferences, we develop a hypothesis to explain the bright state and QPOs in Sec.~\ref{sec:qpo_discussion}.

    \subsection{Nature of the QPOs}
    \label{sec:qpo_discussion}

    The QPOs in the \tess\ light curve have been intermittently present in ground-based photometry of TX~Col \citep{mhlahlo}, but the \textit{TESS} light curve provides several previously unavailable clues concerning their nature. First, they appeared exclusively during epochs of increased mass transfer in a stream-fed geometry. Second, the highest-amplitude QPOs were confined to a relatively narrow range of frequencies between 10-25 cycles d$^{-1}$. The QPOs reported by \citet{sullivan} and \citet{mhlahlo} fall into this range, underscoring that QPOs in this frequency range are a long-term property of TX Col, rather than a unique feature of the bright states observed by \tess. This well-defined frequency range, as well as its long-term repeatability, distinguishes TX~Col's QPOs from those observed during the outbursts of other IPs, such as EX~Hya \citep[][their Fig.~2]{RB90}, and it will be important for future studies to search for evidence of this behavior in other IPs.
    
    Interpreting these characteristics of TX~Col is difficult because we are unaware of previous theoretical predictions that QPOs can be the dominant feature in photometry of stream-fed IPs. Moreover, existing models of QPOs, which were developed for systems with accretion disks, cannot account for the properties of the QPOs in TX Col. \citet{ww02} and \citet{warner04} identify three major classes of QPO-like behaviors in cataclysmic variable stars: (1) dwarf-nova oscillations (DNOs), (2) longer-period DNOs, and (3) QPOs. The two DNO flavors are characterized by high coherence and short periods: $\sim$tens of seconds for the short-period variety, and several minutes for the longer-period DNOs. These properties are obviously incompatible with the incoherence and hours-long timescales of the QPOs in TX~Col. It is clear from \citet{warner04} that QPOs are rather heterogeneous in their observational properties, with low coherence and longer periods than DNOs. Although an excited mode in an accretion disk could produce QPOs with periods of several thousand seconds \citep{warner04}, our observations suggest that TX~Col lacked an accretion disk when the QPOs emerged. We conclude, as did \citet{mhlahlo}, that a different mechanism must be at play in TX~Col.
    
    In the absence of clear-cut theoretical guidance, we propose that during the bright state, the increased ram pressure of the accretion stream enables a torus of diamagnetic blobs to form near the circularization radius,
    \begin{equation}
    \frac{R_c}{a} = (1+q)  \left(\frac{b}{a}\right)^4,
    \end{equation} where $R_c$ is the radius of a circular orbit around the WD whose orbital momentum is equal to that of matter at the $L_1$ point, 
    $a$ is the binary separation, $q = M_2/M_1$, and $b$ is the distance between the inner Lagrangian point and the WD \citep[][Eq.~3]{norton04}. These blobs are predicted by \citet{wk95} to be capable of orbiting the WD $\sim$10 times before being accreted, but the resulting ring-like structure would not interact viscously with itself, since the magnetic timescale is shorter than the viscous timescale \citep{wk95}. We hypothesize that the QPOs are beats between the WD's rotational period and diamagnetic blobs orbiting near $R_c$.
    
    We examined the plausibility of this scenario as follows. The stellar masses of TX~Col are unknown,\footnote{\citet{bt89} presented a circumstantial argument for a very massive ($\sim$1.3M$_{\odot}$~WD) based on the assumption that the full-width-at-zero-intensity of the Balmer emission was produced at the inner rim of the disk. However, this phenomenon could also be explained by emission from the magnetically confined portion of the accretion flow, so while we do not rule out the possibility of a massive WD, we do not think that it is the only possible explanation for the observations in \citet{bt89}.} so we cannot calculate a unique value for the Keplerian frequency at $R_c$ in TX~Col. We can, however, calculate it for a grid of stellar masses and then compare the resulting beat frequency against the observations. The goal of this approach is to identify a characteristic frequency for a quasi-Keplerian orbit at $R_c$, and not to explain the observed range of QPO frequencies.
    
\begin{figure*}
        \centering
        \includegraphics[width=\textwidth]{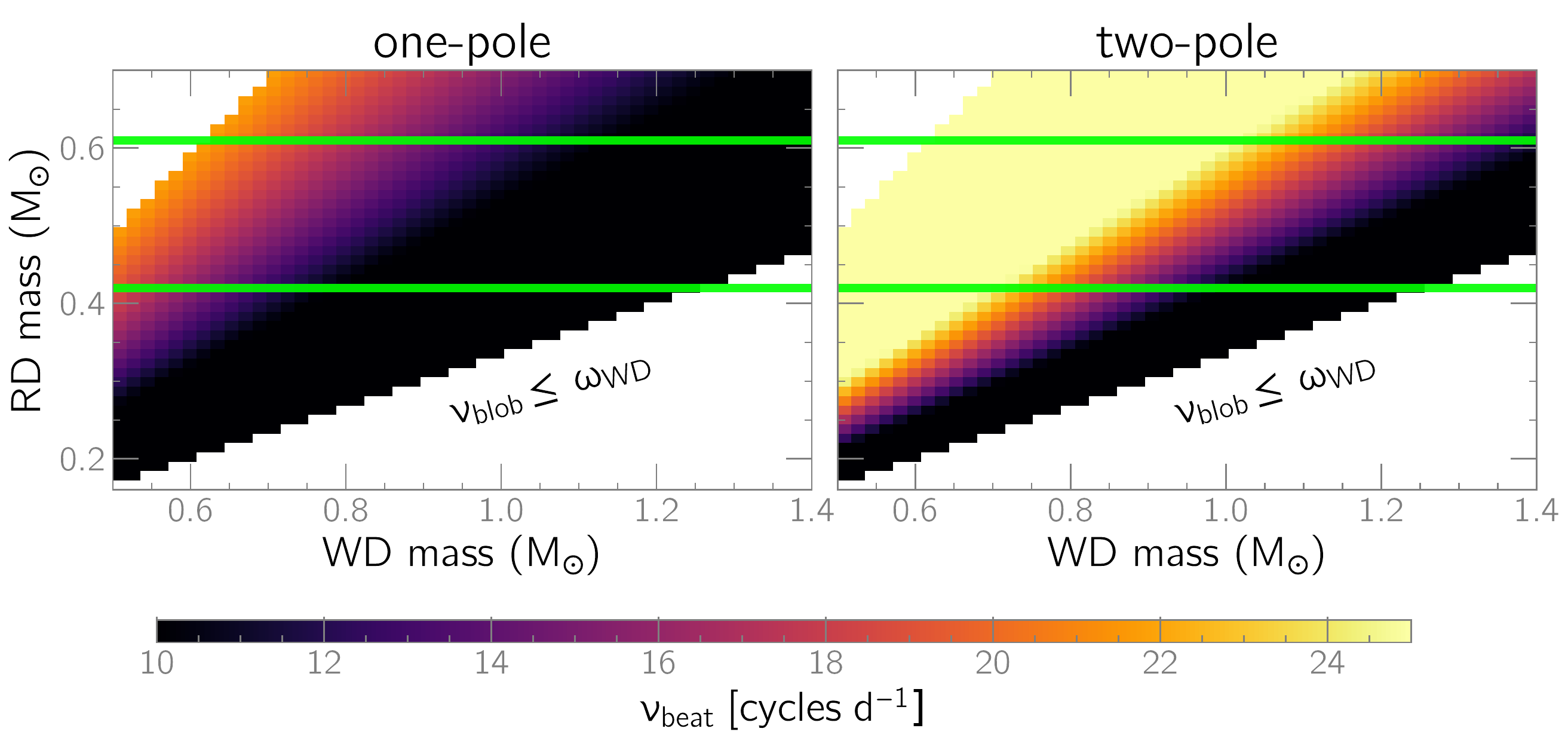}
        \caption{Diagnostic diagram showing the beat frequency of blobs orbiting at the circularization radius, computed for a grid of stellar masses and TX~Col's rotational frequency of $\omega =45$~c~d$^{-1}$. The green horizontal lines indicate $0.42 \leq M_2/M_{\odot} \leq 0.61$, based on the semi-empirical mass-period relation from \citet{warner95}. The limits of the colormap correspond to the observed frequency range of the high-amplitude QPOs. The left panel assumes $\nu_{QPO} = \nu_{blob} - \omega$, which corresponds with a blob interacting with just one magnetic pole per blob orbit; the right panel shows $\nu_{QPO} = 2(\nu_{blob}-\omega)$, wherein an orbiting blob interacts with opposite magnetic poles during opposite halves of its orbit around the WD. Magnetic drag has been neglected, but since it would reduce $\nu_{blob}$, it would consistently reduce the QPO frequency. We conclude that the one-pole model, while plausible, requires a combination of an awkwardly massive secondary and a relatively low-mass WD to produce the full range of QPOs within the observed 10-25~c~d$^{-1}$ corridor from Fig.~\ref{fig:trailed_power}. In contrast, the two-pole model predicts the observed QPO frequencies for reasonable stellar masses at TX~Col's orbital period of 5.7~h. We exclude mass ratios in excess of unity, and we require blobs to orbit faster than the WD's field. }
        \label{fig:QPO_model}
    \end{figure*}

    Accordingly, for a range of stellar masses, we calculated the Keplerian frequency at $R_c$ and found the resulting beat frequency with $\omega$ (Fig.~\ref{fig:QPO_model}). We then compared the predicted beat frequency against the observed QPO frequencies. 
    As an example, for a typical WD mass in a CV of 0.80~M$_{\odot}$ \citep{zorotovic11} and a donor mass\footnote{Conservative limits of $0.42 \leq M_2/M_\odot \leq 0.61$ from Warner's semi-empirical mass-period relation \citep{warner95} are indicated in Fig.~\ref{fig:QPO_model}.} of 0.50~M$_{\odot}$ \citep[typical for $P_{orb} = 5.72$~h; ][]{ck11}, we calculate the Keplerian orbital frequency at $R_c$ to be 59~cycles~d$^{-1}$. When this frequency beats against the WD spin frequency, it yields a QPO frequency of
    \begin{equation}
    \nu_{QPO} = \nu_{blob} - \omega,
    \label{eq_qpo1}
    \end{equation}
    which for the assumed stellar masses is 13~c~d$^{-1}$. This is in reasonable agreement with the observed QPOs near the lower end of the $10-25$~cycles~d$^{-1}$ range. However, since the blobs would be orbiting inside the WD's corotation radius, they would rapidly cross field lines, experiencing a drag force that would slow them below the Keplerian frequency at $R_c$. This decrease in $\nu_{blob}$ would also decrease the QPO frequency, which could present a challenge for explaining the higher-frequency QPOs.

    Eq.~\ref{eq_qpo1} presupposes that a blob interacts with just one magnetic pole as it orbits the WD. However, if a blob survives long enough to revolve around the WD in the WD's rotational rest frame, we would expect it to interact with opposite magnetic poles during opposite halves of the orbit in that frame of reference. Consequently, we would expect to observe QPOs centered on a frequency of \begin{equation}
        \nu_{QPO} = 2(\nu_{blob} - \omega).
        \label{eq_qpo2}
    \end{equation}
    With our example of M$_1$=0.80~M$_{\odot}$ and M$_2$=0.50~M$_{\odot}$, the predicted QPO frequency is $\sim$27~c~d$^{-1}$, which is only marginally higher than the observed QPOs. Magnetic drag would further lower the predicted QPO frequency towards the observed range.
    
    We therefore propose that the QPOs are beats between the rotation of the WD and decaying blob orbits near the circularization radius. A more sophisticated treatment of this problem would account for a range of blob lengths and densities, both of which affect the magnitude of the drag force and therefore $\nu_{blob}$ \citep{wk95}. We speculate that the heterogenous nature of the blobs might be why the QPO frequencies exhibit so much scatter, but since the parameters used to calculate the drag coefficients of the blobs are often poorly constrained, order-of-magnitude estimates, we leave this problem to a future theoretical study.
    
    Although \citet{mhlahlo} and \citet{rawat} also invoked diamagnetic blobs to explain the persistent QPO that they identified in TX~Col, our proposal differs significantly in that we place the blobs much closer to the WD and argue against the presence of a Keplerian accretion disk. In contrast, \citet{mhlahlo} and \citet{rawat} appear to have presumed that the blobs coexist with an accretion disk, with the QPOs being caused by the blobs reprocessing spin-modulated X-rays into the optical. Specifically, \citet{mhlahlo} calculated that blobs with Keplerian periods of $\sim$3000~s could produce the observed QPO frequencies by beating against the WD spin, and they found that the Keplerian period was consistent with an origin in the outer accretion disk. \citet{rawat} also favored the outer accretion disk as the likely location of the blobs. However, it is unclear how the diamagnetic blobs described theoretically by \citet{wk95} could coexist with an accretion disk; presumably, they would collide with the outer rim of the disk shortly after leaving the $L_1$ point. Moreover, the original \citet{wk95} model is an alternative to disk-fed accretion, and it does not include an accretion disk.

    Finally, diamagnetic-blob accretion might also explain the origin of the dips observed during TX~Col's bright states. Some fraction of infalling diamagnetic blobs should be expelled from the WD magnetosphere in a diskless accretion regime \citep{wk95}. Though the ejected matter would not necessarily be expelled from the binary, these episodic ejections would temporarily inhibit accretion, probably producing dips in the light curve.

    \subsection{Disk formation}

    The observation of diamagnetic-blob accretion is widely applicable to the study of IPs because it offers a possible explanation as to how accretion disks can form in IPs with large magnetospheres. This is a particularly salient consideration for IPs that experience large, prolonged variations in $\dot{M}$, because these systems can alternate between disk-fed and diskless accretion geometries. Using FO~Aqr as an example, \citet{HL17} present a strong theoretical argument that some IPs can contain Keplerian accretion disks during epochs of high $\dot{M}$, but if $\dot{M}$ decreases for a prolonged interval, the magnetospheric radius can exceed the disk's circularization radius, causing the disk to dissipate altogether.
    
    The reestablishment of a Keplerian disk in such a system has been the subject of debate, because no disk can be present if the magnetospheric radius is larger than the circularization radius. To resolve this conundrum, \citet{hellier02} postulated that diamagnetic-blob accretion at a sufficiently high $\dot{M}$ might cause blobs to screen each other from the WD's magnetic field, pile up, and spread into a disk. It was beyond the scope of that paper to model this scenario in detail, and we are not aware of any subsequent work that has done so. However, if the blobs could survive for at least their viscous timescale, it is natural to expect that they would interact and develop into a Keplerian disk \citep{hellier02}. Thus, we speculate that at an even higher mass-transfer rate in TX Col, the torus of blobs might be able to form a Keplerian disk near the circularization radius. In such a case, we might expect to observe a cessation of the large-amplitude QPOs and a transition to the $\omega$-dominated power spectrum of a disk-fed system \citep{fw99}.

    \section{Conclusion}
    
    Our analysis of the \textit{TESS} observations of TX~Col from Cycles~1~and~3 showed diskless accretion during the system's normal brightness state. When TX~Col brightened, large-amplitude QPOs overwhelmed the periodic variability observed at lower accretion rates. The QPO-dominated light curve persisted for $\sim$4 weeks in both cycles. The strongest QPOs were confined to a well-defined range of frequencies ($\sim10-25$~cycles~d$^{-1}$) and are consistent with those identified in previously reported ground-based photometry \citep{mhlahlo}. Consequently, they are a long-term characteristic of TX~Col. Our analysis strongly suggests that the presence of the QPOs is predicated upon an enhanced accretion rate.
    
    Existing models do not offer a satisfactory explanation for the QPOs in TX~Col. Instead, we propose that the QPOs are beats between the WD's rotational frequency and the orbital frequencies of diamagnetic blobs that orbit inside the WD's magnetosphere. We speculate that TX~Col was therefore on the precipice of forming a Keplerian accretion disk. However, the diamagnetic blob model suffers from a lack of directly testable theoretical predictions, and we encourage additional work in this area.

    \software{ {\tt astropy} \citep{astropy}, {\tt lightkurve} \citep{lightkurve} }

    \acknowledgements
    
    We thank the anonymous referee for an expeditious and well-reasoned report, particular their valuable criticism of our speculation about EX~Hya in the original preprint. The referee's insights persuaded us to drop the corresponding subsection from the final manuscript.

    PS and CL acknowledge support from NSF grant AST-1514737. M.R.K. acknowledges support from the ERC under the European Union’s Horizon 2020 research and innovation programme (grant agreement No. 715051; Spiders). 
    \clearpage

    \bibliography{bib.bib}

\begin{thebibliography}{}
\expandafter\ifx\csname natexlab\endcsname\relax\def\natexlab#1{#1}\fi
\providecommand{\url}[1]{\href{#1}{#1}}
\providecommand{\dodoi}[1]{doi:~\href{http://doi.org/#1}{\nolinkurl{#1}}}
\providecommand{\doeprint}[1]{\href{http://ascl.net/#1}{\nolinkurl{http://ascl.net/#1}}}
\providecommand{\doarXiv}[1]{\href{https://arxiv.org/abs/#1}{\nolinkurl{https://arxiv.org/abs/#1}}}

\bibitem[{{Astropy Collaboration} {et~al.}(2013){Astropy Collaboration},
  {Robitaille}, {Tollerud}, {Greenfield}, {Droettboom}, {Bray}, {Aldcroft},
  {Davis}, {Ginsburg}, {Price-Whelan}, {Kerzendorf}, {Conley}, {Crighton},
  {Barbary}, {Muna}, {Ferguson}, {Grollier}, {Parikh}, {Nair}, {Unther},
  {Deil}, {Woillez}, {Conseil}, {Kramer}, {Turner}, {Singer}, {Fox}, {Weaver},
  {Zabalza}, {Edwards}, {Azalee Bostroem}, {Burke}, {Casey}, {Crawford},
  {Dencheva}, {Ely}, {Jenness}, {Labrie}, {Lim}, {Pierfederici}, {Pontzen},
  {Ptak}, {Refsdal}, {Servillat}, \& {Streicher}}]{astropy}
{Astropy Collaboration}, {Robitaille}, T.~P., {Tollerud}, E.~J., {et~al.} 2013,
  \aap, 558, A33, \dodoi{10.1051/0004-6361/201322068}

\bibitem[{{Bailer-Jones} {et~al.}(2021){Bailer-Jones}, {Rybizki}, {Fouesneau},
  {Demleitner}, \& {Andrae}}]{BJ21}
{Bailer-Jones}, C.~A.~L., {Rybizki}, J., {Fouesneau}, M., {Demleitner}, M., \&
  {Andrae}, R. 2021, \aj, 161, 147, \dodoi{10.3847/1538-3881/abd806}

\bibitem[{{Buckley} \& {Schwarzenberg-Czerny}(1992)}]{ex_hya_dip}
{Buckley}, D.~A.~H., \& {Schwarzenberg-Czerny}, A. 1992, in Astronomical
  Society of the Pacific Conference Series, Vol.~29, Cataclysmic Variable
  Stars, ed. N.~{Vogt}, 354

\bibitem[{{Buckley} \& {Sullivan}(1992)}]{buckley92}
{Buckley}, D.~A.~H., \& {Sullivan}, D.~J. 1992, Astronomical Society of the
  Pacific Conference Series, Vol.~29, {The Remarkable Period Changes in the
  Intemediate Polar TX Columbae}, ed. N.~{Vogt}, 387

\bibitem[{{Buckley} \& {Tuohy}(1989)}]{bt89}
{Buckley}, D.~A.~H., \& {Tuohy}, I.~R. 1989, \apj, 344, 376,
  \dodoi{10.1086/167806}

\bibitem[{{Ferrario} \& {Wickramasinghe}(1999)}]{fw99}
{Ferrario}, L., \& {Wickramasinghe}, D.~T. 1999, \mnras, 309, 517,
  \dodoi{10.1046/j.1365-8711.1999.02860.x}

\bibitem[{{Hameury} \& {Lasota}(2017)}]{HL17}
{Hameury}, J.~M., \& {Lasota}, J.~P. 2017, \aap, 606, A7,
  \dodoi{10.1051/0004-6361/201731226}

\bibitem[{{Hellier} \& {Beardmore}(2002)}]{hellier02}
{Hellier}, C., \& {Beardmore}, A.~P. 2002, \mnras, 331, 407,
  \dodoi{10.1046/j.1365-8711.2002.05199.x}

\bibitem[{{Kennedy} {et~al.}(2016){Kennedy}, {Garnavich}, {Breedt}, {Marsh},
  {G{\"a}nsicke}, {Steeghs}, {Szkody}, \& {Dai}}]{kennedy}
{Kennedy}, M.~R., {Garnavich}, P., {Breedt}, E., {et~al.} 2016, \mnras, 459,
  3622, \dodoi{10.1093/mnras/stw834}

\bibitem[{{King} \& {Wynn}(1999)}]{kw99}
{King}, A.~R., \& {Wynn}, G.~A. 1999, \mnras, 310, 203,
  \dodoi{10.1046/j.1365-8711.1999.02974.x}

\bibitem[{{Knigge} {et~al.}(2011){Knigge}, {Baraffe}, \& {Patterson}}]{ck11}
{Knigge}, C., {Baraffe}, I., \& {Patterson}, J. 2011, \apjs, 194, 28,
  \dodoi{10.1088/0067-0049/194/2/28}

\bibitem[{{Kochanek} {et~al.}(2017){Kochanek}, {Shappee}, {Stanek}, {Holoien},
  {Thompson}, {Prieto}, {Dong}, {Shields}, {Will}, {Britt}, {Perzanowski}, \&
  {Pojma{\'n}ski}}]{kochanek}
{Kochanek}, C.~S., {Shappee}, B.~J., {Stanek}, K.~Z., {et~al.} 2017, \pasp,
  129, 104502, \dodoi{10.1088/1538-3873/aa80d9}

\bibitem[{{Lightkurve Collaboration} {et~al.}(2018){Lightkurve Collaboration},
  {Cardoso}, {Hedges}, {Gully-Santiago}, {Saunders}, {Cody}, {Barclay}, {Hall},
  {Sagear}, {Turtelboom}, {Zhang}, {Tzanidakis}, {Mighell}, {Coughlin}, {Bell},
  {Berta-Thompson}, {Williams}, {Dotson}, \& {Barentsen}}]{lightkurve}
{Lightkurve Collaboration}, {Cardoso}, J. V. d.~M., {Hedges}, C., {et~al.}
  2018, {Lightkurve: Kepler and TESS time series analysis in Python}.
\newblock \doeprint{1812.013}

\bibitem[{{Littlefield} {et~al.}(2020){Littlefield}, {Garnavich}, {Kennedy},
  {Patterson}, {Kemp}, {Stiller}, {Hambsch}, {Heras}, {Myers}, {Stone},
  {Sj{\"o}berg}, {Dvorak}, {Nelson}, {Popov}, {Bonnardeau}, {Vanmunster}, {de
  Miguel}, {Alton}, {Harris}, {Cook}, {Graham}, {Brincat}, {Lane}, {Foster},
  {Pickard}, {Sabo}, {Vietje}, {Lemay}, {Briol}, {Krumm}, {Dadighat}, {Goff},
  {Solomon}, {Padovan}, {Bolt}, {Kardasis}, {Deback{\`e}re}, {Thrush}, {Stein},
  {Walter}, {Coulter}, {Tsehmeystrenko}, {Gout}, {Lewin}, {Galdies},
  {Fernandez}, {Walker}, {Boardman}, \& {Pellett}}]{littlefield}
{Littlefield}, C., {Garnavich}, P., {Kennedy}, M.~R., {et~al.} 2020, \apj, 896,
  116, \dodoi{10.3847/1538-4357/ab9197}

\bibitem[{{Mhlahlo} {et~al.}(2007){Mhlahlo}, {Buckley}, {Dhillon}, {Potter},
  {Warner}, {Woudt}, {Bolt}, {McCormick}, {Rea}, {Sullivan}, \&
  {Velhuis}}]{mhlahlo}
{Mhlahlo}, N., {Buckley}, D.~A.~H., {Dhillon}, V.~S., {et~al.} 2007, \mnras,
  380, 133, \dodoi{10.1111/j.1365-2966.2007.12003.x}

\bibitem[{{Norton} {et~al.}(1997){Norton}, {Hellier}, {Beardmore}, {Wheatley},
  {Osborne}, \& {Taylor}}]{norton}
{Norton}, A.~J., {Hellier}, C., {Beardmore}, A.~P., {et~al.} 1997, \mnras, 289,
  362, \dodoi{10.1093/mnras/289.2.362}

\bibitem[{{Norton} {et~al.}(2004){Norton}, {Wynn}, \& {Somerscales}}]{norton04}
{Norton}, A.~J., {Wynn}, G.~A., \& {Somerscales}, R.~V. 2004, \apj, 614, 349,
  \dodoi{10.1086/423333}

\bibitem[{{Patterson}(1994)}]{patterson94}
{Patterson}, J. 1994, \pasp, 106, 209, \dodoi{10.1086/133375}

\bibitem[{{Rawat} {et~al.}(2021){Rawat}, {Pandey}, \& {Joshi}}]{rawat}
{Rawat}, N., {Pandey}, J.~C., \& {Joshi}, A. 2021, arXiv e-prints,
  arXiv:2104.06944.
\newblock \doarXiv{2104.06944}

\bibitem[{{Reinsch} \& {Beuermann}(1990)}]{RB90}
{Reinsch}, K., \& {Beuermann}, K. 1990, \aap, 240, 360

\bibitem[{{Retter} {et~al.}(2005){Retter}, {Liu}, \& {Bos}}]{retter}
{Retter}, A., {Liu}, A., \& {Bos}, M. 2005, Astrophysics and Space Science
  Library, Vol. 332, {Evidence for Large Superhumps in TX Col and V4742 Sgr},
  ed. E.~M. {Sion}, S.~{Vennes}, \& H.~L. {Shipman}, 251--259,
  \dodoi{10.1007/1-4020-3725-2_25}

\bibitem[{{Scaringi} {et~al.}(2017){Scaringi}, {Maccarone}, {D'Angelo},
  {Knigge}, \& {Groot}}]{scaringi17}
{Scaringi}, S., {Maccarone}, T.~J., {D'Angelo}, C., {Knigge}, C., \& {Groot},
  P.~J. 2017, \nat, 552, 210, \dodoi{10.1038/nature24653}

\bibitem[{{Shappee} {et~al.}(2014){Shappee}, {Prieto}, {Grupe}, {Kochanek},
  {Stanek}, {De Rosa}, {Mathur}, {Zu}, {Peterson}, {Pogge}, {Komossa}, {Im},
  {Jencson}, {Holoien}, {Basu}, {Beacom}, {Szczygie{\l}}, {Brimacombe},
  {Adams}, {Campillay}, {Choi}, {Contreras}, {Dietrich}, {Dubberley},
  {Elphick}, {Foale}, {Giustini}, {Gonzalez}, {Hawkins}, {Howell}, {Hsiao},
  {Koss}, {Leighly}, {Morrell}, {Mudd}, {Mullins}, {Nugent}, {Parrent},
  {Phillips}, {Pojmanski}, {Rosing}, {Ross}, {Sand}, {Terndrup}, {Valenti},
  {Walker}, \& {Yoon}}]{shappee}
{Shappee}, B.~J., {Prieto}, J.~L., {Grupe}, D., {et~al.} 2014, \apj, 788, 48,
  \dodoi{10.1088/0004-637X/788/1/48}

\bibitem[{{Sullivan} {et~al.}(1995){Sullivan}, {Buckley}, \&
  {Thomas}}]{sullivan}
{Sullivan}, D.~J., {Buckley}, D.~A.~H., \& {Thomas}, C. 1995, Astronomical
  Society of the Pacific Conference Series, Vol.~85, {Recent Multi-site
  Observations of TX Columbae}, ed. D.~A.~H. {Buckley} \& B.~{Warner}, 512

\bibitem[{{Szkody} {et~al.}(2017){Szkody}, {Mukadam}, {Toloza}, {G{\"a}nsicke},
  {Dai}, {Pala}, {Waagen}, {Godon}, \& {Sion}}]{rzleo}
{Szkody}, P., {Mukadam}, A.~S., {Toloza}, O., {et~al.} 2017, \aj, 153, 123,
  \dodoi{10.3847/1538-3881/aa5c88}

\bibitem[{{Tuohy} {et~al.}(1986){Tuohy}, {Buckley}, {Remillard}, {Bradt}, \&
  {Schwartz}}]{tuohy86}
{Tuohy}, I.~R., {Buckley}, D.~A.~H., {Remillard}, R.~A., {Bradt}, H.~V., \&
  {Schwartz}, D.~A. 1986, \apj, 311, 275, \dodoi{10.1086/164770}

\bibitem[{{Warner}(1986)}]{warner86}
{Warner}, B. 1986, \mnras, 219, 347, \dodoi{10.1093/mnras/219.2.347}

\bibitem[{{Warner}(1995)}]{warner95}
---. 1995, {Cataclysmic variable stars}, Vol.~28

\bibitem[{{Warner}(2004)}]{warner04}
---. 2004, \pasp, 116, 115, \dodoi{10.1086/381742}

\bibitem[{{Warner} \& {Woudt}(2002)}]{ww02}
{Warner}, B., \& {Woudt}, P.~A. 2002, \mnras, 335, 84,
  \dodoi{10.1046/j.1365-8711.2002.05596.x}

\bibitem[{{Wynn} \& {King}(1995)}]{wk95}
{Wynn}, G.~A., \& {King}, A.~R. 1995, \mnras, 275, 9,
  \dodoi{10.1093/mnras/275.1.9}

\bibitem[{{Zorotovic} {et~al.}(2011){Zorotovic}, {Schreiber}, \&
  {G{\"a}nsicke}}]{zorotovic11}
{Zorotovic}, M., {Schreiber}, M.~R., \& {G{\"a}nsicke}, B.~T. 2011, \aap, 536,
  A42, \dodoi{10.1051/0004-6361/201116626}

\end{thebibliography}

\end{document}